\documentclass[aip,apl,preprint]{revtex4-1}

\usepackage{graphicx, abbrevs}

\begin{document}
Copyright (2012) American Institute of Physics. This article may be downloaded for personal use only. Any other use requires prior permission of the author and the American Institute of Physics.

The following article appeared in Appl. Phys. Lett. 101, 013115 (2012); doi: 10.1063/1.4733292 and may be found at http://link.aip.org/link/?APL/101/013115
\newpage

\title{Optical energy optimization at the nanoscale by near-field interference}

\newabbrev\FDTD{Finite Difference Time Domain (FDTD)}[FDTD]
\newabbrev\EMF{electromagnetic (EM)}[EM]
\newabbrev\AOS{helicity dependent all-optical switching (HD-AOS)}[HD-AOS]

\author{Benny Koene}
\email{b.koene@science.ru.nl}
\author{Matteo Savoini}
\author{Alexey V. Kimel}
\author{Andrei Kirilyuk}
\author{Theo Rasing}
\affiliation{Radboud University Nijmegen, Institute for Molecules and Materials, Heyendaalseweg 135, 6525~AJ Nijmegen, The Netherlands}

\date{\today}

\begin{abstract}
Employing plasmonic antennas for subdiffraction focusing of light on recording media requires to take into account the complete structure of the medium, including dielectric cover layers. We find, with Finite Difference Time Domain simulations, that optical energy transfer to the magnetic recording layer is most efficient for an off-resonant antenna. Furthermore we show that the focal spot in the magnetic film is well below the diffraction limit, making nanoscale all-optical magnetic data recording achievable.
\end{abstract}


\maketitle 

Due to their ability to confine light below the diffraction limit and the accompanying large intensity enhancement \cite{Mikhailovsky2003,Mikhailovsky2004,Celebrano2009,Novotny2011,Biagioni2012}, optical antennas are ideal for increasing the efficiency of light-matter interactions. These properties make optical antennas not only useful for applications in the field of photovoltaics, nonlinear optics, and quantum optics \cite{Bharadwaj2009}, but also in data storage technologies. 

In Heat Assisted Magnetic Recording (HAMR), plasmonic structures are used to heat nanoscale spots, such that their magnetization can then be reversed with a smaller magnetic field, increasing the potential data storage density \cite{Challener2009,Stipe2010}. With HAMR an external magnetic field is still necessary. However, exploiting \AOS magnetic domains can be switched reversibly with circularly polarized femtosecond laser pulses in the absence of any external magnetic field, making all-optical data storage and retrieval possible \cite{Kimel2005,Stanciu2007,Vahaplar2009}.

Although the \AOS process is very fast, the reported sizes of the switched domains are of several microns, which is not appealing for industry, where the actual bit size is already well below 100~nm. The plasmonic structures used in HAMR can however not be implemented for \AOS as they do not preserve the light polarization. To maintain the circular polarization state in the near field we will make use of cross antennas. With this type of antennas large enhancement factors have been demonstrated in the antenna gap\cite{Biagioni2009,Biagioni2009a}. However, as the actual recording media are usually protected by a dielectric or metallic capping layer of several tens of nanometers, this will severely affect the focusing abilities and subsequent potential data densities for \AOS.

Here we demonstrate that it is possible to deliver energy to the magnetic layer more efficiently by exploiting the near-field interference between the excitation light and the re-emitted light \cite{Barnard2011}. In particular we show that due to this interference an off-resonant antenna delivers more energy to a distant plane as compared to a resonant one. Evaluating the spot size and field enhancement that can be obtained inside the magnetic film shows that we have a gain in energy together with a sub-diffraction sized spot, even in the presence of a capping layer. These two quantities are of direct relevance for \AOS.

\begin{figure}
	\includegraphics{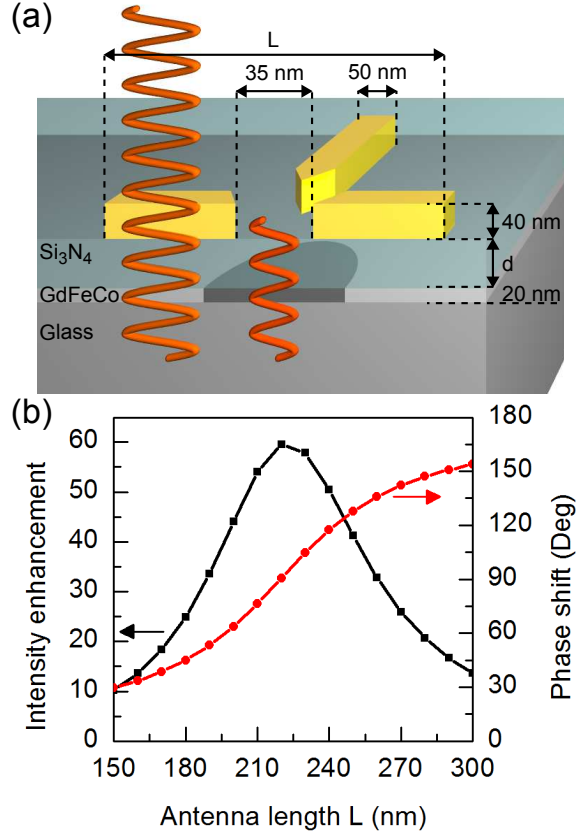}
 	\caption{\label{Sketch}Crossed dipole antenna. (a) A sketch of the considered situation. An incoming wave(left) excites the antenna and interferes with the re-emitted wave(right). In the GdFeCo layer this results in a nanoscale reversed magnetic domain(dark cylinder) due to HD-AOS. (b) Intensity enhancement and phase shift as a function of the antenna length. Both are recorded in the center of the antenna gap and in the presence of a 60~nm $\mathrm{Si}_{3}\mathrm{N}_{4}$ layer.}
\end{figure}

The structure we consider is as follows: a glass substrate (dielectric constant \cite{Note1} $\epsilon = 2.11$) \cite{Palik1997}, a thin film of magnetic material, in our case 20~nm of GdFeCo ($\epsilon = -1.15+28.56\mathrm{i}$) \cite{Hendren2003}, protected by a capping layer of $\mathrm{Si}_{3}\mathrm{N}_{4}$ ($\epsilon = 4$) \cite{Palik1997}. The sample is similar to the samples typically used in \AOS experiments \cite{Stanciu2007,Vahaplar2009}. In those experiments a capping layer of 60~nm is used to optimize the magneto-optical signal. On top of the last layer we place the cross antenna structure. We use \FDTD simulations \cite{Lumerical} to calculate the \EMF fields at different positions in the structure presented in Fig. \ref{Sketch}(a). A circularly polarized plane wave at a wavelength of 800~nm is used to excite the antenna. In the simulations we assume typical dimensions for real cross antennas: a thickness of 40~nm, an arm width of 50~nm, and a gap size of 35~nm\cite{Biagioni2011}. The total length of two opposite arms including the gap is varied from 110~nm to 300~nm. The antenna is made of gold ($\epsilon = -24.11+1.49\mathrm{i}$) \cite{Johnson1972}. In the \FDTD simulations a non-uniform meshing is used with the smallest mesh cells 1$\times$1$\times$1~nm at the position of the antenna, extending to the GdFeCo film directly underneath the antenna.

In general, when plasmonic antennas are considered, the attention is focused on the intensity enhancement. This intensity enhancement is calculated by normalizing the field intensities obtained in the presence of the antenna structure with the results of a simulation without the antenna. Fig. \ref{Sketch}(b) shows this intensity enhancement in the antenna gap when the antenna length is varied and a 60~nm thick $\mathrm{Si}_{3}\mathrm{N}_{4}$ capping layer is used. However, the intensity enhancement as defined earlier does not take into account the difference in energy delivery to the magnetic thin film due to a change in the capping layer thickness, while this capping layer, which is effectively a cavity, will influence the energy delivery. As \AOS depends on the polarization of the \EMF fields we would like to have a measure for the degree of circular polarization as well. For these reasons we will from here on use the following figure of merit (FOM)\cite{Biagioni2009,Biagioni2009a} defined by
\begin{equation}\label{EQ:FOM}
	\mathrm{FOM} = IC^2,
\end{equation}
where $I$ is the intensity and $C$ is the degree of circular polarization given by
\begin{equation}\label{EQ:C}
	\mathrm{C} = \frac{2 E_x E_y \sin(\delta_x - \delta_y)}{I}.
\end{equation}
Here $E_i$ and $\delta_i$ are respectively the amplitude and phase of the field component $i$. For $I$ we will take the field intensity normalized to the source intensity. Note that the source intensity is equall for all simulations and thus it will take into account the difference in energy delivery due to the capping layer thickness. We would like to emphasize that we do not use the intensity enhancement here, which we can get if we normalize to the intensity profile obtained without antenna in the same plane. 

\begin{figure}
	\includegraphics{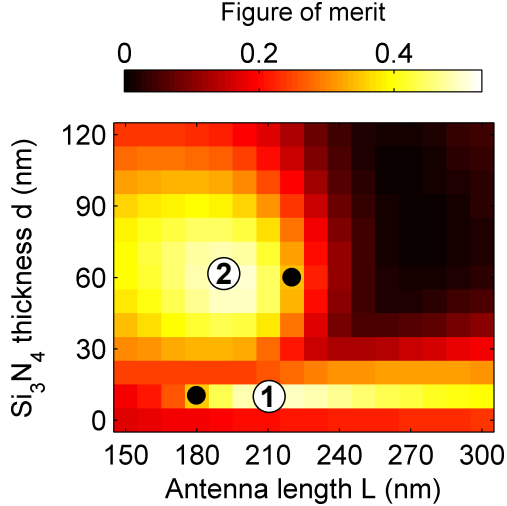}
 	\caption{\label{Layerthickness}FOM as function of the  antenna length L and the $\mathrm{Si}_{3}\mathrm{N}_{4}$ layer thickness d. The FOM is recorded at half-height of the GdFeCo layer directly below the center of the antenna. Two maxima are observed, indicated by the numbers one and two. The black dots indicate the lengths at which the antenna is resonant for the two optimal $\mathrm{Si}_{3}\mathrm{N}_{4}$ thicknesses.}
\end{figure}

In Fig. \ref{Layerthickness} the FOM inside the GdFeCo is shown as a function of the capping layer thickness and the antenna length. Two maxima can be observed, the first maximum can be found for a dielectric layer thickness of 10~nm and an antenna length of 210~nm. The second maximum is reached for a capping layer thickness of 60~nm and an antenna length of 190~nm. For both capping layer thicknesses this maximum does not coincide with the antenna resonant length as defined in the antenna gap and indicated by the black dots in Fig. \ref{Layerthickness}. Note that when the capping layer thickness is changed the antenna resonant length and quality can be different as well. Close to the metallic magnetic film the FOM inside the antenna gap will be reduced significantly due to the presence of an extra loss channel formed by the lossy plasmon modes bound to the GdFeCo surface. This directly explains the equal amplitude of the two maxima despite the difference in distance.  

\begin{figure}
	\includegraphics{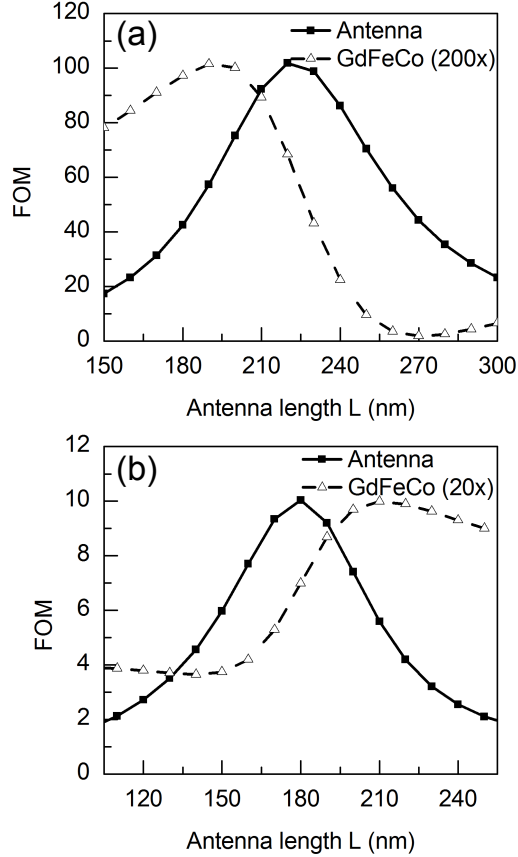}
 	\caption{\label{Interference}FOM in the GdFeCo (open symbols) as function of antenna length. The solid symbols show the FOM in the center of the antenna gap. In (a) a 60~nm $\mathrm{Si}_{3}\mathrm{N}_{4}$ layer is present while in (b) this layer is 10~nm thick.}
\end{figure}

To explain the origin of the discrepancy between the antenna lengths at which the FOM in the GdFeCo reaches a maximum and the antenna resonant length, we will concentrate on the capping layer thicknesses at which these maxima occur. In Fig. \ref{Interference} the FOM inside the GdFeCo (open symbols) is shown together with the FOM in the center of the antenna gap (solid symbols) at a fixed $\mathrm{Si}_{3}\mathrm{N}_{4}$ thickness of 60~nm(a) and 10~nm(b). These figures show a clear difference between the behavior of the FOM with antenna length in the plane of the antenna and inside the GdFeCo. The maximum intensity in the GdFeCo layer occurs for 10~nm (60~nm) of $\mathrm{Si}_{3}\mathrm{N}_{4}$ for an antenna longer (shorter) than the resonant length.

The observed behavior can be explained by interference between the re-emitted light from the antenna and the excitation light. Like a harmonic oscillator, a resonance in intensity goes together with a phase shift for the re-emitted light. The phase shift in the center of the antenna gap as function of the antenna length is shown in Fig. \ref{Sketch}(b) for the 60~nm thick capping layer. In the center of the antenna gap no interference can be observed due to the large difference in intensity of the two light sources. From Fig. \ref{Sketch}(b) it can be seen that the light in the antenna gap is 60 times more intense than the exciting light. However, the exponential decay of the plasmon fields causes this large difference in intensity to disappear fast with distance. This results in comparable intensities for the two light beams in the GdFeCo layer, 60~nm below the antenna, leading to interference effects. The increase in intensity towards the plasmon resonance combined with the relative small phase difference for antennas shorter than the resonant length makes the maximum intensity in the active layer to occur for an antenna that is a few tens of nanometers shorter than a resonant antenna. For antennas longer than resonant ones, the large phase shift for the re-emitted light results in destructive interference, which explains the minimum in Fig. \ref{Interference}(a) for an antenna length of 270~nm.

If we now compare the above situation with the case of only a 10~nm capping layer, Fig. \ref{Interference}(b), than we can observe a few differences. First, in Fig. \ref{Interference}(b) there is no complete destructive interference visible. This could indicate that the two \EMF waves that interfere do not have the same amplitude. The other difference is that while in the 60~nm case we have a maximum for an antenna that is shorter than the resonant one, for the 10~nm case we have in the same situation a minimum. We ascribe this effective phase shift of roughly $\pi$ for one of the two waves to the interplay between the plasmon modes of the antenna and the GdFeCo film.

\begin{figure*}
	\includegraphics{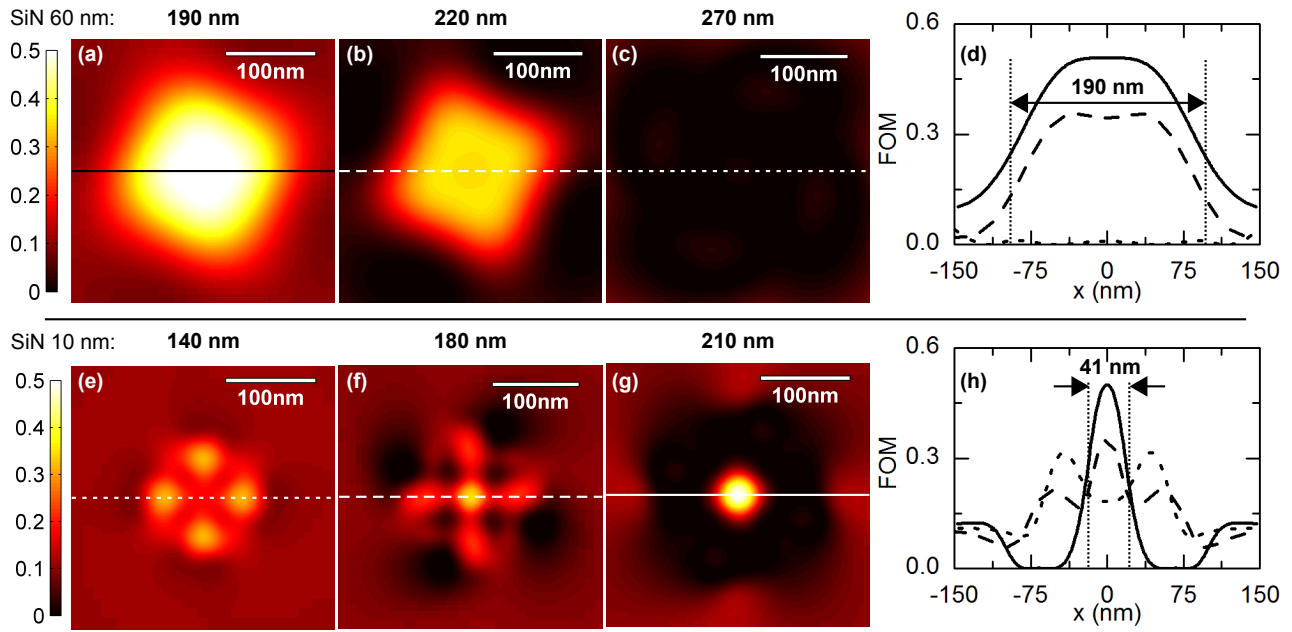}
 	\caption{\label{areaplots}FOM distribution in a 2D plane at half height of the GdFeCo layer in case of a 60~nm (a)-(c) and a 10~nm (e)-(g) dielectric layer for different antenna lengths. In (d) and (h) cross-sections through the center of respectively (a)-(c) and (e)-(g) are shown. The FWHM of the central spot is indicated for the cross-section of 190~nm (d) and 210~nm (h) antenna length.}
\end{figure*}

All previous simulation results were collected in a single point only. However, for the purpose of magnetic switching it is interesting to know the field patterns resulting from the interference in the plane of the magnetic thin film. These interference patterns give an indication about the size of the magnetic domain that is switched by using the optical antennas as considered in this paper. The obtained interference patterns for the 60~nm thick capping layer are shown in Figs. \ref{areaplots}(a)-(c) for antenna lengths of respectively 190~nm (constructive interference), 220~nm (plasmon resonance) and 270~nm (destructive interference). For the antenna with a length of 190~nm a spot size with a full width at half maximum (FWHM) of 190~nm is found, as is clearly shown in Fig. \ref{areaplots}(d). Note that we have a constant FOM over a width of 90~nm. If we calculate the maximum intensity enhancement in the GdFeCo for this antenna we find a value of 2.6.

Figs. \ref{areaplots}(e)-(h) shows the same information as Figs. \ref{areaplots}(a)-(d) but now for a 10~nm capping layer. With this thickness we find an even smaller spot size with a FWHM of 41~nm. This sub diffraction area is substantially smaller than the spot size of a light beam focused with conventional objectives (typical dimensions with an immersion oil objective are of approximately 350~nm using 800~nm light). We would like to mention that considering only the field intensity $I$ would give a spot size FWHM of 52~nm. Hence we have here an advantage due to the dependence of the AOS on the helicity of light. The maximum field enhancement in the 41~nm spot is 3.7. This would mean that the known switching treshold of $2.6~\mathrm{mJ}/\mathrm{cm}^{2}$ for a GdFeCo sample without any structure on top \cite{Khorsand2012}, will reduce to $0.7~\mathrm{mJ}/\mathrm{cm}^{2}$ in the case that nanoantennas are used.

Although low, the intensity enhancement we find is in the same order of magnitude of others reported for devices aimed for similar applications (as for example reported in Fig. 2(b) of Ref.~\citenum{Challener2009}). In our case the low value is caused by the quickly vanishing nature of the near fields together with the presence of a capping layer while for example in Ref.~\citenum{Challener2009} there is a less efficient coupling due to the high refractive index surrounding the antenna. For the purpose of \AOS this small enhancement factor is not a problem as the main reason to use optical antennas here is to bring down switching to the nanoscale. 

In this paper we considered a plasmonic structure that can be made with the current state of the art nanofabrication technologies \cite{Biagioni2011}. With progress in the fabrication process smaller antenna gaps will be attainable. This would make it possible to have smaller spot sizes and larger enhancement factors. Considering data storage technology, in which a writing head (antenna) moves over the medium, we performed some simulations with a 7~nm air slit between the antenna and the capping layer as well. Except in the antenna length no noticeable changes where observed.

To conclude, we have shown that besides the direct field enhancement and field confinement related to plasmonic structures, near-field interference effects can be of similar importance when optimizing the energy delivery to the near field. As a relevant example we considered a metallic magnetic thin film protected by a dielectric capping layer with cross-antennas on top. The antenna length at which the maximum energy is delivered to the thin film differs from the resonant antenna length. This difference can be attributed to the interference between the excitation light and the re-emitted light by the antenna. With the antennas on top of a 10~nm thick capping layer and using 800~nm light, we were able to obtain a sub diffraction spot with a maximum intensity enhancement of 3.7 and a FWHM of 41~nm inside the magnetic material, which is comparable to the bit size in present day storage technology.

\begin{acknowledgments}
This work was supported by de Nederlandse Organisatie voor Wetenschappelijk Onderzoek (NWO) and de Stichting voor Fundamenteel Onderzoek der Materie (FOM). The authors would like to thank Paolo Biagioni, Marco Finazzi and Lamberto Du\`{o} for fruitful discussions.
\end{acknowledgments}


%

\end{document}